\documentstyle[preprint,prl,aps]{revtex}
\begin{document}
\draft
\title{Anomalous magnetophotoluminescence as a result of level
repulsion in arrays of quantum dots.}
\author{E. V. Tsiper$^1$, P. D. Wang$^{2*}$, J. L. Merz$^2$, A. L.
Efros$^1$, S. Fafard$^{3\dagger}$, D. Leonard$^{3\ddagger}$, and
P. \nolinebreak M. \nolinebreak Petroff$^3$}
\address{$^1$Department of Physics, University of Utah, Salt Lake City,
UT 84112}
\address{$^2$Department of Electrical Engineering, University of Notre
Dame, Notre Dame, IN 46556}
\address{$^3$Center for Quantized Electronic Structures (QUEST),
Materials Department, and Department of Electrical and Computer
Engineering University of California, Santa Barbara, CA 93106}

\date{\today}
\maketitle
\newpage

\begin{abstract}

Selectively excited photoluminescence (SPL) of an array of
self-organized In$_{0.5}$Ga$_{0.5}$As quantum dots has been measured
in a magnetic field up to 11T.  Anomalous magnetic field sensitivity
of the SPL spectra has been observed under conditions for which the
regular photoluminescence spectra is insensitive to the magnetic field
due to large inhomogeneous broadening.  The anomalous sensitivity is
interpreted in terms of the repulsion of excited levels of the dots in
a random potential.  A theory presented to describe this phenomena is
in excellent agreement with the experimental data.  The data estimated
the correlation in the positions of excited levels of the dots to be
94\%.  The magnetic field dependence allows the determination of the
reduced cyclotron effective mass in a dot.  For our sample we have
obtained $m_em_h/(m_e+m_h)=0.034m_0$.

\end{abstract}

An array of self-organized quantum dots (QDs) is a unique system
consisting of very small atomic-like objects each with a few energy
levels\cite{Leonard94b,Moison94,Marzin94,Grundmann95}.
The studies of this system have shown the possibility to attain
three-dimensional confinement of carriers within QDs.
Such quantum dots are formed in highly strained semiconductor
heterostructures by what is known as Stranski-Krastanow growth, where
growth starts two-dimensionally, but after a certain critical
thickness is reached, islands are formed spontaneously, and a thin
wetting layer is left under the islands.  In this process, the growth
is interrupted immediately after the formation of the islands and
before strain relaxation and misfit dislocations occur.  Such in-situ
formation of 0D quantum dots results in high quality defect-free
materials.  In addition, the coherent islanding and strain effects can
produce QDs with a size uniformity within $\pm$10\% which is very
promising for 0D quantum devices where the sharper density of states
is exploited.

Photoluminescence (PL) spectrum of such an array has a broad line
which is supposed to be mainly due to inhomogeneous
broadening\cite{fafard94,wang97}.  Photoluminescence excitation (PLE)
and selectively-excited photoluminescence (SPL) reveal a fine
structure.  This fine structure has been interpreted by one of
us\cite{rr} as a result of splitting of the excited levels in quantum
dots due to violation of cylindrical symmetry of the dots by a random
potential.  Though the splitting is much less than the PL linewidth,
it can be observed because of the effect of repulsion of energy levels
inside a dot.  Here we give an experimental proof of this point and a
detailed study of the level repulsion in quantum dots by applying
magnetic field which affects the splitting of excited levels.

We found an anomalous sensitivity of the SPL to a magnetic field under
conditions for which the regular photoluminescence spectra is
insensitive to the magnetic field due to large inhomogeneous
broadening.  We show that this sensitivity is a direct result of the
level repulsion.  A preliminary discussion of this effect has been
given earlier\cite{rr_talks}.

The dot layer studied here is pseudomorphically grown by MBE on (100)
GaAs substrate, and the QDs are formed by the coherent relaxation into
islands of a few monolayers (ML) of In$_{0.5}$Ga$_{0.5}$As between
GaAs buffer and cap layers.  The actual amount of indium incorporated
in the dots can differ due to the complex dynamics of the adatoms
during island formation.  The growth and QD structural details have
been reported earlier\cite{Leon95}.

Fig.~1(a) shows PLE and regular PL spectra.  Regular PL reveals a
broad line with the FWHM=57meV.  PLE spectra consists of
significantly narrower lines.  PL spectrum is red-shifted with respect
to PLE by about 80 meV.  This shift occurs since the PLE experiment
detects the light emitted only by the levels below the level excited
by the pumping light.

The SPL spectra for different excitation energies are presented in
Fig.~1(b).  The complicated character of the SPL spectra were analyzed
earlier\cite{rr}.  For our purposes it is important that at some
energy of excitation ($E_{ex}=1.3672$eV) the SPL spectrum shows two
symmetric peaks.  At larger and smaller $E_{ex}$ these peaks become
asymmetric.

For an intuitive picture it is helpful to assume that each dot has two
optically active excited levels $E_\pm$ relatively close to each other
(see Fig.~2).  The dots are isolated from each other, so the light is
emitted from the same dot it is absorbed to.  The dots are excited
into $E_+$ or $E_-$ and, after thermalization, emit light from $E_0$.
The red shift mentioned above originates from the difference
$E_\pm-E_0$.  In some dots the excitation energy $\hbar\omega_{ex}$
may be close either to $E_-$ or to $E_+$.  These two types of dots can
be considered as two different subsets.  These subsets should give two
peaks in the SPL curve if we assume the correlation between the
positions of $E_\pm$ and $E_0$.  If this correlation is of such kind
that the dots with larger $E_\pm$ have, in general, larger $E_0$, the
excitation of $E_+$ will cause the lower peak in SPL.  This is a
natural proposal and it corresponds to our experimental data (see
Fig.~1(b)).  At some energy of excitation there are the same amounts
of dots in each subset.  However, if the energy of excitation is
shifted upward, the number of the dots which have the lowest excited
level at this energy is smaller, so the intensity of the higher-energy
peak decreases.  Correspondingly, the low-energy peak disappears as
the excitation energy decreases.

In our interpretation two close optically active excited levels
originate from the doubly degenerate lowest excited level of a
cylindrically symmetric dot.  These levels are split by a random
potential which may include a deviation of a dot shape from
cylindrical symmetry.  It is crucial that both the distance between
the peaks and their width, as well as the linewidth of the
nonselective PL have the same origin.  They are determined by the
random potential which splits degenerate levels and shifts randomly
all energy levels in QDs.

At the excitation energy $E_{ex}=1.3672$eV, which gives two symmetric
peaks in the SPL, we have studied the magnetic field dependence of the
SPL spectrum.  The results presented in Fig.~3 show anomalous
sensitivity to the magnetic field.  The relative intensity of the dip
between two peaks decreases by 11\% in the magnetic field of 2T.  Note
that the wide line of the regular (non-selective) PL is insensitive to
a magnetic field up to $\sim$10T\cite{Wang96}.

We show below that the two-level structure of SPL at zero field and
its anomalous sensitivity to the magnetic field are both the results
of the level repulsion.

For an axially symmetric dot two degenerate wave functions with an
angular momentum $|m|$ have the form $\Psi_{m\pm}({\bf r})=\psi_m(r)
e^{\pm im\phi}$, where $\psi_m$ can be chosen real. For the first
excited state $m=1$.  The positions of energy levels $\epsilon_\pm$,
split and shifted by random potential and by magnetic field, can be
obtained as the eigenvalues of the secular matrix

\begin{equation}
\delta H=\left[
\begin{tabular}{cc}
$u+\frac{\hbar\omega}{2}$, & $x+iy$\\
$x-iy$, & $u-\frac{\hbar\omega}{2}$
\end{tabular}
\right],
\label{matr}
\end{equation}
where the matrix elements $u$ and $x+iy$ take random values in each
quantum dot and are given by\cite{rr,RE}:
\begin{eqnarray}
u&=&\int d^3r\ V({\bf r})\psi_1^2({\bf r}),\\
x+iy&=&\int d^3r\ V({\bf r})\psi_1^2({\bf r})
   e^{2i\phi}.
\end{eqnarray}
Here $V({\bf r})$ is arbitrary Gaussian random potential.  It can be
caused by alloy fluctuations.  In principle, the same description is
also valid in the case when the cylindrical symmetry is violated by
strain field or shape of the dots\cite{rr}.

The values $u$, $x$, and $y$ are real independent Gaussian random
variables with equal standard deviations $\sigma_1$.  The eigenvalues
of $\delta H$ are $\epsilon_\pm=u\pm\Delta$, where $\Delta$ is the
splitting of the excited level given by

\begin{equation}
\Delta=\sqrt{\Delta_0^2+(\hbar\omega/2)^2},
\label{delta}
\end{equation}
$\Delta_0=\sqrt{x^2+y^2}$ being the splitting in the absence of the
magnetic field.  Magnetic field provides an extra splitting of the
excited level, which for an axially-symmetric dot is equal to the
cyclotron energy $\hbar\omega$.

The distribution function for $\Delta$ is easy to calculate:

\begin{equation}
F(\Delta)=\frac{2\Delta}{\sigma_1^2}
  \exp\left(-\frac{\Delta^2-(\hbar\omega/2)^2}{\sigma_1^2}\right)
  \theta\left(\Delta^2-(\hbar\omega/2)^2\right),
\label{F}
\end{equation}
where $\theta$ is the step function.  At zero magnetic field $F(0)=0$
due to the level repulsion.  In a finite field $F(\Delta)=0$ in the
region $|\Delta|<\hbar\omega/2$ since the splitting cannot be less
than $\hbar\omega$.  Now we show that PLE and SPL are closely related
to the function $F(\Delta)$.

To find lineshape for both PLE and SPL, one should calculate intensity
as a function of two frequencies $I(E_{ex},E_{det})$.  The PLE and SPL
lineshapes can be obtained from this function by fixing corresponding
variables.  We assume that matrix elements are energy independent, so
the intensity is proportional to the distribution function
$P(\epsilon_0,\epsilon_1)$ where $\epsilon_0$ and $\epsilon_1$ are the
energies of the lowest and the next excited states respectively.  To
simplify the notation we measure energies from their average values,
$\epsilon_\alpha=E_\alpha-\overline{E_\alpha}$.

The lowest state $E_0$ with the wave function $\psi_0(r)$ is also
shifted by random potential by

\begin{equation}
\epsilon_0=\int d^3r\ V({\bf r})\psi_0^2({\bf r}),
\end{equation}
The shift $\epsilon_0$ is also a Gaussian random variable which is
statistically independent of $x$ and $y$, however, in general, it is
correlated with $u$.

The general expression for $P(\epsilon_0,\epsilon_1)$ has the form:

\begin{eqnarray}
P(\epsilon_0,\epsilon_1)&=&\sum_\pm\int\!\int\!\int dudxdy
  \ \delta(\epsilon_1-u\mp\sqrt{x^2+y^2+(\hbar\omega/2)^2})\nonumber\\
&\times& G_2(\epsilon_0,u;\sigma_0,\sigma_1,\rho)
  G_1(x;\sigma_1/\sqrt{2})G_1(y;\sigma_1/\sqrt{2})
\label{Pi}
\end{eqnarray}
Here $\sigma_0$ is the dispersion of $\epsilon_0$.

The matrix element $u$ determines the overall shift of the excited
level due to the random potential.  $G_1(\epsilon;\sigma)$ is the
normal distribution,
$G_1(\epsilon;\sigma)=\exp(-\epsilon^2/2\sigma^2)/\sigma\sqrt{2\pi}$.

The general form of the two-variable normal distribution of the
variables $\epsilon_0$ and $u$ which takes into account the
correlation between them is

\begin{equation}
G_2(\epsilon_0,u;\sigma_0,\sigma_1,\rho)=
  \frac{1}{2\pi\sigma_0\sigma_1\sqrt{1-\rho^2}}
  \exp\left\{-\frac{1}{2(1-\rho^2)}
  \left[\frac{\epsilon_0^2}{\sigma_0^2}
    -2\rho\frac{\epsilon_0u}{\sigma_0\sigma_1}
    +\frac{u^2}{\sigma_1^2}\right]\right\},
\label{G2}
\end{equation}
Here $\rho$ is the correlation coefficient, $|\rho|\leq1$.

Taking integrals in Eq.~(\ref{Pi}) one obtains
\begin{equation}
P(\epsilon_0,\epsilon_1)=G_1(\epsilon_0;\sigma_0)
  D_\rho(\epsilon_1-\epsilon_0\rho\frac{\sigma_1}{\sigma_0};\sigma_1),
\label{P}
\end{equation}
where the function $D_\rho(\epsilon;\sigma)$ is defined by

\begin{eqnarray}
D_\rho(\epsilon;\sigma)&=&\frac{1}{\sigma}\sum_{(\pm\epsilon)}
\left\{
  \frac{\sqrt{\mu-1}}{\mu\sqrt{\pi}}
    \exp\left(-\frac{(\hbar\omega/2-\epsilon)^2}{(\mu-1)\sigma^2}\right)+
\right.\nonumber\\
&&\left.
\frac{\epsilon}{\sigma\mu^{3/2}}
    \exp\left(\frac{\mu(\hbar\omega/2)^2-\epsilon^2}{\mu\sigma^2}\right)
    \left[1-\text{erf}\left(
      \frac{\mu\hbar\omega/2-\epsilon}{\sigma\sqrt{\mu(\mu-1)}}
  \right)\right]\right\},
\label{D}
\end{eqnarray}
where $\mu=3-2\rho^2$.  The function $P(\epsilon_0,\epsilon_1)$ as
determined by Eqs.~(\ref{P}), (\ref{D}) is the generalization of the
function introduced in \cite{rr} for zero magnetic field.
Eq.~(\ref{D}) yields Eq.~(\ref{F}) with $\epsilon=\Delta$ when
$\rho\rightarrow1$.

The effect of level repulsion manifests itself in SPL and in PLE in
full scale if the overall shift of $E_\pm$ is proportional to the
shift of $E_0$.  This means $\rho=1$.  In this case we get

\begin{eqnarray}
P(\epsilon_0,\epsilon_1)=\frac{1}{\sigma_0\sqrt{2\pi}}
  \exp\left(-\frac{\epsilon_0^2}{2\sigma_0^2}\right)
  F\left(\epsilon_1-\epsilon_0\frac{\sigma_0}{\sigma_1}\right),
\label{rho1}
\end{eqnarray}
where $F$ is the distribution function of splittings given by
Eq.~(\ref{F}).

The PLE lineshape can be obtained from Eq.~(\ref{rho1}) by fixing the
detection energy $\epsilon_0$.  One can see that it is just given by
the function $F$.  To get SPL lineshape one should fix $\epsilon_1$.
The lineshape of the SPL also reproduces the features of the function
$F$.  Because of the level repulsion both SPL and PLE are zero when
$\epsilon_1=\epsilon_0\sigma_1/\sigma_0$.  In the magnetic field they
are zero in the range
$|\epsilon_1-\epsilon_0\sigma_1/\sigma_0|<\hbar\omega/2$.

If the fluctuations of the lowest and the next excited states were
uncorrelated, the structure will be substantially smeared.  However,
the correlation occurs to be strong.  In the model of alloy disorder
the correlation coefficient $\rho$ is calculated to be
$\rho=0.795$\cite{rr}.  The experimental value we obtain below is
$\rho=0.945$.  This coefficient describes the correlation in
positions of the lowest excited level $E_0$ and the next split
excited level $E_\pm$.  The value of $\rho$ close to 1 suggests that a
substantial part of the fluctuations originates from the change of the
size of the dots which shifts energy levels proportionally.  That is
why such a tiny effect as the repulsion of split levels makes SPL
spectra very sensitive to a relatively weak magnetic field.

Fig.~4 shows the relative peak position and the relative valley
intensity as a function of magnetic field.  Solid lines represent the
best fit to the experimental data with Eqs.~(\ref{P}), (\ref{D}).  The
fit has been performed in the following way.  The parameters $\rho$,
$\sigma_0$, and $\sigma_1$ can be determined from zero-field data. 
Two dimensionless parameters, $\rho$ and $\sigma_0/\sigma_1$, are
obtained from the relative depth of the dip in SPL and the shift of
PLE maxima with detection energy as provided by Eq.~(\ref{P}).  This
gives $\rho=0.945$ and
$\sigma_0/\sigma_1=\rho/(dE_{PLE}/d\hbar\omega_{det})=1.189$.  The
absolute value of $\sigma_1$ can be found from the relative peak
position at zero field, which gives $\sigma_1=21.4$meV and
$\sigma_0=25.5$meV.  To check the consistency, we can obtain
$\sigma_0$ from the width of the non-selective PL in Fig.~1(a).
Assuming the Gaussian shape of PL, the FWHM=57meV gives
$\sigma_0=57/(2\sqrt{2\log2})=24.2$meV. 

With these parameters given, the magnetic-field dependence of both the
relative depth of the dip and the relative peak position depend only
on the effective mass.  The value $m_{\text{eff}}=0.034m_0$ gives
perfect fit for both quantities as shown in Fig.~4.

The value of the effective mass we obtained can be understood by
taking into account that the cyclotron frequency that enters
Eq.~(\ref{matr}) contains the reduced effective mass for the electron
and hole, $m_{\text{eff}}=m_em_h/(m_e+m_h)$.  If we assume $m_e=m_h$,
our result would imply $m_e=0.068m_0$.  On the other hand, if we take
for the effective mass of the electron the value for bulk
Ga$_{0.5}$In$_{0.5}$As, $m_e=0.045$\cite{LB}, the value of
$m_{\text{eff}}=0.034m_0$ would give for the mass of the hole
$m_h=0.14m_0$.  The small value of the hole mass is natural taking
into account the two-dimensional nature of the QDs.  On the other
hand, the effects of strain and confinement are believed to increase
the effective mass of the carriers with respect to the bulk
material\cite{Cusack96,People90}.

In conclusion, we have observed anomalous sensitivity of SPL to the
magnetic field low enough to affect the regular PL spectrum.  We
interpret such sensitivity in terms of the repulsion of energy levels
in QDs caused by a random potential.  We present a theory which
describes the phenomenon and shows excellent agreement with the
experimental data.  The data allows to determine the correlation in
the positions of excited levels the reduced cyclotron effective mass
for the carriers in the dots.

\section*{Acknowledgments}
A.L.E. acknowledges the support of UCSB, subcontract No. KK3017 of QUEST.

\begin{figure}
\caption{PL and PLE (a) and SPL (b) spectra of InGaAs QDs with
different excitation and detection energies.}
\label{fig.1}
\end{figure}

\begin{figure}
\caption{An intuitive picture of relevant energy levels in QDs.  Note
that excitation into the higher level gives lower-energy feature in
SPL}
\label{fig.2}
\end{figure}

\begin{figure}
\caption{SPL spectra under different magnetic fields.  The excitation
energy is $\hbar\omega_{ex}$=1.3673eV.}
\label{fig.3}
\end{figure}

\begin{figure}
\caption{The relative valley intensity (squares) and peak position
(triangles) as a function of magnetic field.  The solid lines
represent the best fit by Eqs.~(9), (10).  The parameters used are
$\rho=0.945$, $\sigma_0=25.5$meV, $\sigma_1=21.4$meV, and
$m_{\text{eff}}=0.034m_0$.}
\label{fig.4}
\end{figure}

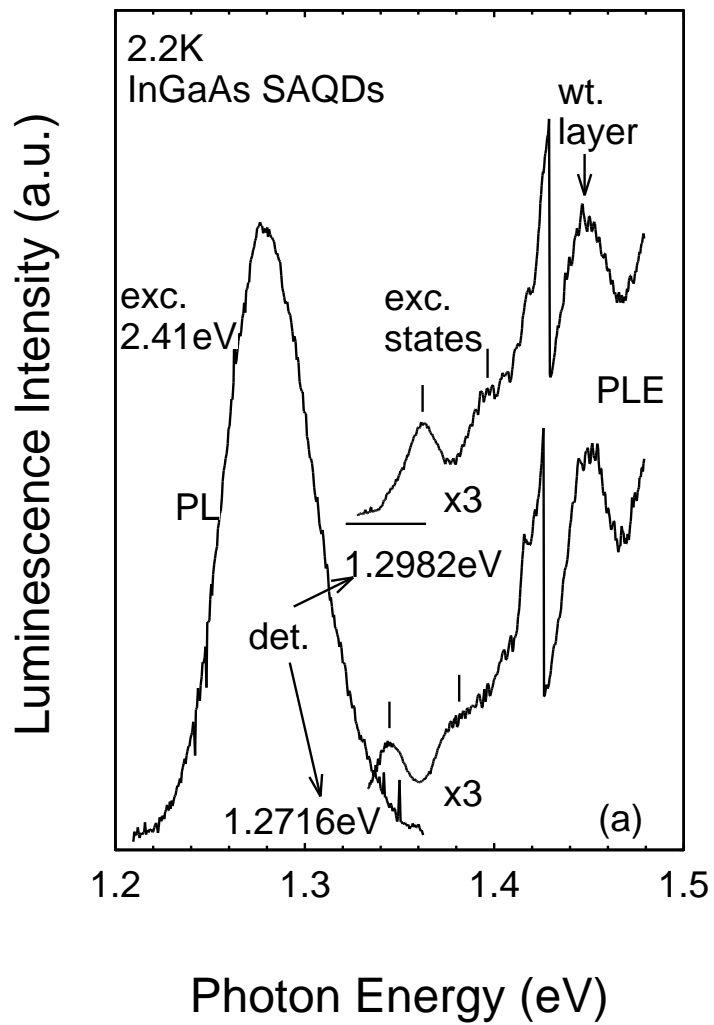
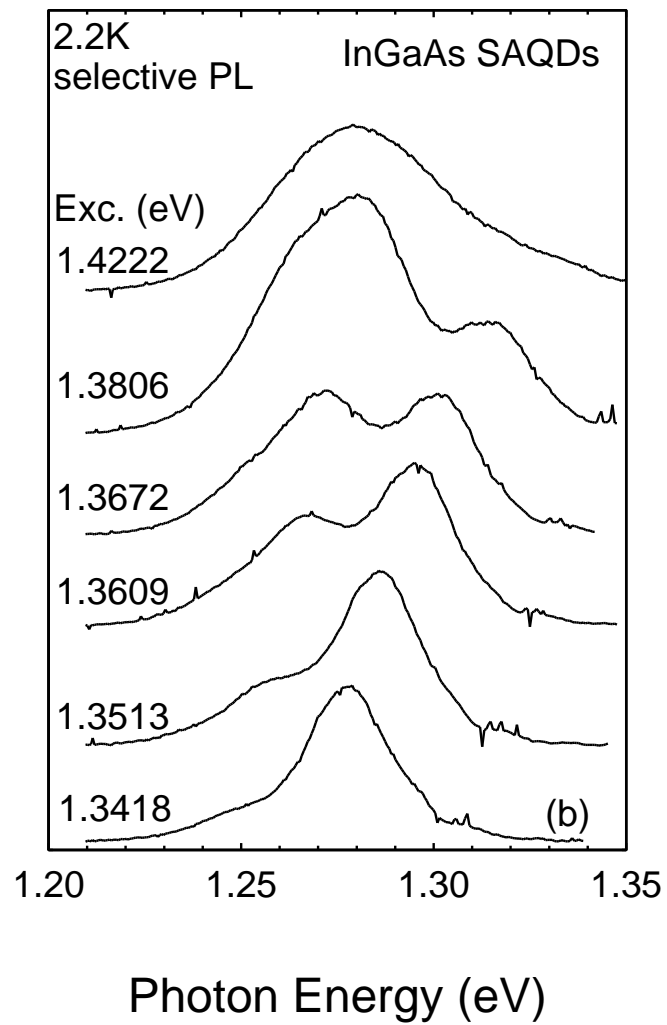
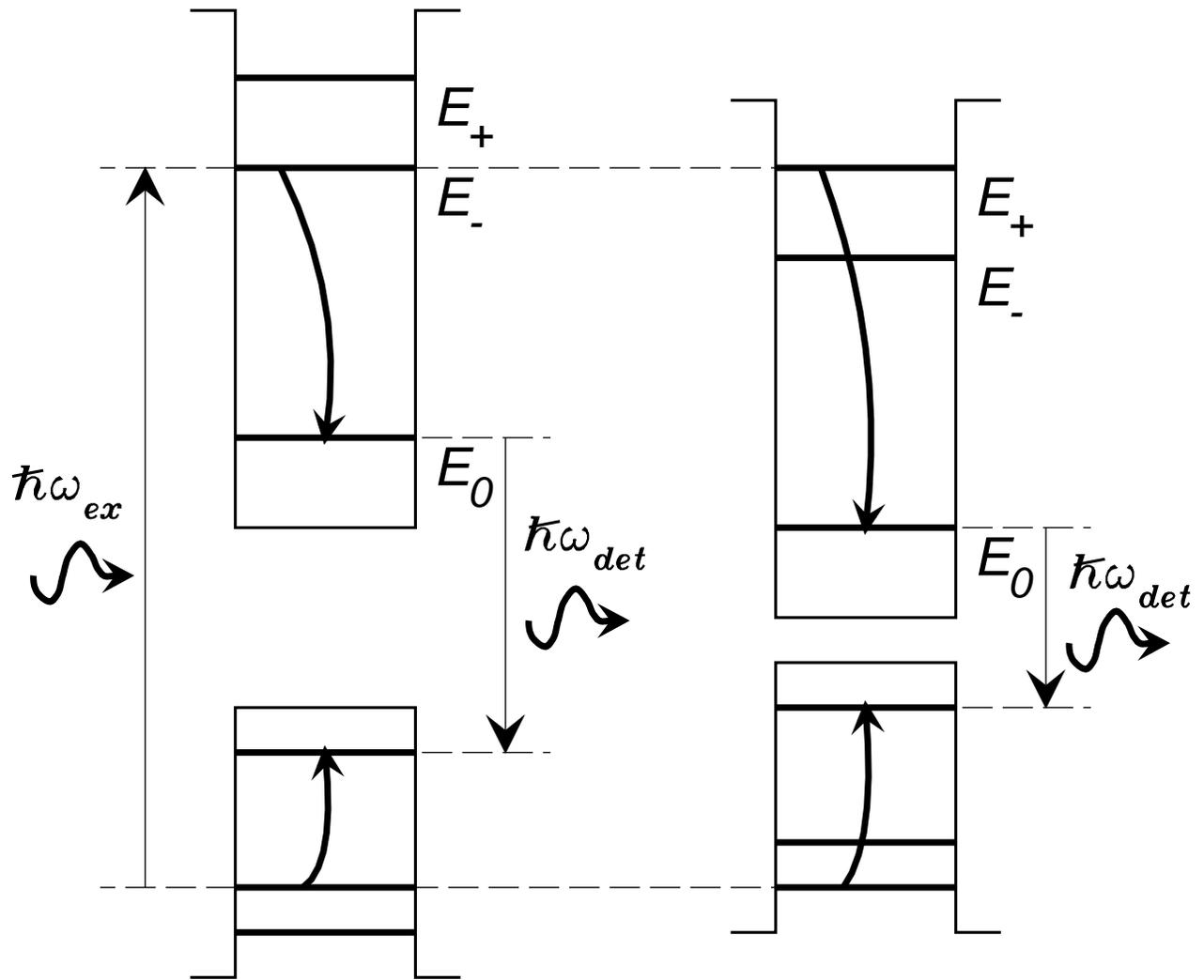
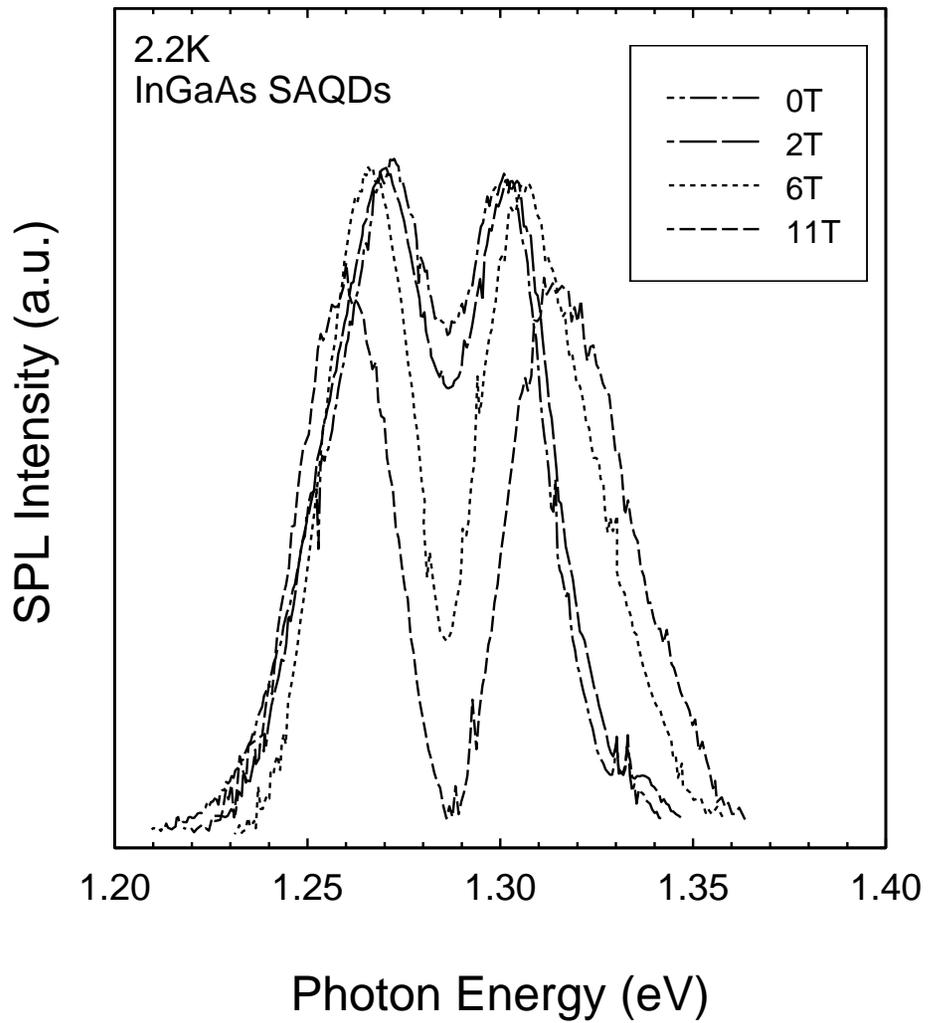
\end{document}